 \documentclass[preprint,12pt,3p]{elsarticle}

\usepackage[utf8]{inputenc}
\usepackage[T1]{fontenc}
\usepackage{amssymb}
\usepackage{amsthm}
\usepackage{amsmath}
\usepackage{color}
\usepackage{dsfont}
\usepackage{mathrsfs}  
\usepackage{mathtools, cuted}
\usepackage{caption}
\usepackage{subcaption}
\usepackage{enumerate}

\newcommand{\vac}{\mathrm{vac}}

\begin{document}

\begin{frontmatter}

\title{Cluster dynamics modelling of materials: a new hybrid deterministic/stochastic coupling approach.}

\author[label1,label2]{Pierre Terrier\corref{cor1}}
\address[label1]{Université Paris-Est, CERMICS (ENPC), INRIA, F-77455 Marne-la-Vallée, France}
\address[label2]{CEA, DEN, Service de Recherches de Métallurgie Physique, UPSay, F-91191 Gif-sur-Yvette, France}

\cortext[cor1]{Corresponding author}
\ead{pierre.terrier@enpc.fr}

\author[label2]{Manuel Athènes}

\author[label2]{Thomas Jourdan}

\author[label3]{Gilles Adjanor}
\address[label3]{EDF R\&D, Matériaux et Mécanique des Composants, Les Renardières, F-77250 Moret sur Loing, France}

\author[label1]{Gabriel Stoltz}

\begin{abstract}
Deterministic simulations of the rate equations governing cluster dynamics in materials are limited by the number of equations to integrate. Stochastic simulations are limited by the high frequency of certain events. We propose a coupling method combining deterministic and stochastic approaches. It allows handling different time scale phenomena for cluster dynamics. This method, based on a splitting of the dynamics, is generic and we highlight two different hybrid deterministic/stochastic methods. These coupling schemes are highly parallelizable and specifically designed to treat large size cluster problems. The proof of concept is made on a simple model of vacancy clustering under thermal ageing.
\end{abstract}

\begin{keyword}
Cluster dynamics, Fokker-Planck equation, Langevin dynamics, Markov process, stochastic simulations
\end{keyword}

\end{frontmatter}

% \linenumbers

\section{Introduction}
\label{sec:intro}

Long time scale phenomena arise in the evolution of the microstructure of materials under thermal ageing or irradiation. To simulate such events (nucleation, formation of precipitates, growth of bubbles etc.) one needs efficient methods that are able to handle systems with different time scales. Monte Carlo methods, such as kinetic Monte Carlo~\cite{voter2007introduction,caturla2000comparative,domain2004simulation}, give accurate results but may be limited to short time simulations when frequent events occur.

Mean-field techniques such as cluster dynamics have been used with success to get around this issue~\cite{ortiz2007he,duparc2002microstructure,jourdan2014efficient}. The modelling of the microstructure is approximated by considering only the defect concentrations, whose evolutions are determined by a system of ordinary differential equations (ODE), called rate equation. Nevertheless two main difficulties occur with cluster dynamics. First, since there is one rate equation per cluster type, the number of equations might become very large (clusters might contain up to millions of atoms or defects) so that solving such a system of ODE becomes computationally prohibitive as the cluster sizes increase. Moreover such systems of ODE are generally stiff, \textit{i.e.} the typical time scale for some reactions is very large while it may be very small for others.

Several methods and approximations have been proposed to solve these equations. Deterministic ones include grouping methods where rate equations are gathered into classes \cite{kiritani1973analysis, golubov2001grouping}, and a Fokker-Planck approach where rate equations for large size clusters are approximated by a Fokker-Planck equation \cite{wolfer1977theory}. Recent developments of the Fokker-Planck approach \cite{jourdan2014efficient,jourdan2016accurate} have proven to be really efficient when only one or two types of defect are considered. They are however strongly limited by the dimensionality of the system. To our knowledge no system with three types of defects or more have been treated with a deterministic approach for large size clusters.

The mean-field formalism of cluster dynamics has been related to purely stochastic approaches such as the well known Stochastic Simulation Algorithm (SSA) introduced by Gillespie \cite{gillespie2000chemical}. Marian \textit{et al.} recently proposed a stochastic implementation of the rate theory approach \cite{marian2011stochastic}. This method is intended to take into account complex clusters containing different species (defects, atoms, etc.). Nevertheless, in stiff systems where certain reactions occur frequently, the efficiency of the computations is still limited by the discrepancies in the time scales.

Attempts have been made to take advantage of both deterministic and stochastic methods. Hybrid deterministic/stochastic algorithms have been proposed. Rate equations are used for small size clusters, while large size ones are handled with a stochastic treatment. In particular, Surh \textit{et al.} \cite{surh2004master} approximate the evolution of large size clusters with a Fokker-Planck equation and use a Langevin dynamics to propagate stochastic particles when clusters reach a certain size. While, to our knowledge, this method is the first one to use such a coupling for cluster dynamics, it has an important limitation. As stochastic particles representing clusters of a certain size are emitted one by one at each time step, the time step should be small to increase the number of particles, and hence reduce the statistical noise. On the contrary, if one wants to rapidly reach large time scales, the time step should be large.

We propose in this work an alternative way to couple deterministic and stochastic simulations. After a brief presentation of the physical model, we introduce in Section~\ref{sec:2} a first splitting between the vacancy concentration and the remainder of the distribution. Hence, when the vacancy concentration is fixed, the cluster dynamics becomes linear. We take advantage of this linearity to introduce a further decomposition of the dynamics, this time between small and large size clusters. We describe a generic version of the algorithm based on these two splittings. This generic method allows us to design several coupling methods depending on the way the subproblems are solved. In particular we introduce in Section~\ref{sec:large} two stochastic methods for computing the evolution of large size clusters, one based on the Markov process related to the rate equations and the other on a Fokker-Planck approximation. In Section~\ref{sec:Cvac}, we present different ways of computing the vacancy concentration. The numerical results presented in Section~\ref{sec:results} confirm the validity of our approximation and the accuracy of the method. We finally discuss improvements/extensions of the method and future work in Section~\ref{sec:conclusion}.

\section{Model description and main algorithm}
\label{sec:2}

We study vacancy clustering during ageing with the model system described in~\cite{ovcharenko2003gmic++}. The chosen model is simple but can be enriched with additional sink/source terms, mobile clusters of size two or greater, etc.

\subsection{Rate equations}
\label{sec:model}

The cluster dynamics approach is used to describe the evolution of cluster size concentration $(C_\vac,C_2,C_3,\cdots)$ where $C_\vac$ is the vacancy concentration and $C_n$ is the concentration of a cluster of size $n$. It is a set of rate equations governing the time evolution of each concentration. We assume that only mono-vacancies are mobile. Therefore the rate equation for the concentration $C_n$ of an immobile cluster of size $n \ge 2$ is
\begin{equation}
    \frac{dC_n}{dt} = \beta_{n-1} C_{n-1} C_\vac - \left( \beta_n C_\vac + \alpha_n \right) C_n + \alpha_{n+1} C_{n+1}.
    \label{eq:EPn}
\end{equation}
In this equation, $\beta_n$ is the absorption rate and $\alpha_n$ the emission rate. These rates take the form~\cite{ovcharenko2003gmic++}
\begin{equation}
\left\{
\begin{aligned}
    \beta_n &= \beta_0 n^{1/3}, \quad n \ge 1 \\[0.2cm]
    \alpha_n &= \alpha_0 n^{1/3}\exp\left( -\frac{E_\vac^{\rm b}(n)}{k_{\rm B} T}\right),
\end{aligned}
\right.
\end{equation}
where $\alpha_0 = \beta_0 = (48\pi^2/V_\mathrm{at}^2)^{1/3} D_\vac$, with $V_\mathrm{at}$ the atomic volume and $D_\vac$ the diffusion coefficient of vacancies. The term $E_\vac^b(n)$ represents the binding energy of a vacancy with a cluster of size $n$:
\begin{equation}
E_\vac^{\rm b}(n) = E_\vac^{\rm f} - \frac{2\gamma V_\mathrm{at}}{r(n)},
\end{equation}
where $\gamma$ is a surface energy, $r$ is the radius of void given by $r(n) = (3nV_\mathrm{at}/4\pi)^{1/3}$ and $E_\vac^{\rm f}$ is the vacancy formation energy. The rate equation for $C_\vac$ is given by:
\begin{equation}
\frac{dC_\vac}{dt} =  - 2\beta_1 C_\vac^2 - \sum_{n\ge 2} \beta_n C_n C_\vac + \sum_{n\ge 2} \alpha_n C_n + \alpha_2 C_2.
\label{eq:EPv}
\end{equation}
The latter equation is obtained by requiring that cluster dynamics preserves the total quantity of matter $\mathcal{Q}_\mathrm{tot}$:
\begin{equation}
\frac{d\mathcal{Q}_\mathrm{tot}}{dt} = \frac{d}{dt}\left( C_\vac + \sum_{n\ge2} nC_n \right) = 0.
\label{eq:Qtot}
\end{equation}
Initial conditions are given by 
\begin{equation}
C_\vac(0) = C_\mathrm{init} \quad\text{and}\quad C_n(0) = 0, \quad n\ge 2,
\label{eq:init} 
\end{equation}
where $C_\mathrm{init}$ is the quenched-in vacancy concentration.

\subsection{Splitting of the dynamics}
\label{sec:maths}
Solving the set of rate equations~\eqref{eq:EPn}--\eqref{eq:EPv} when large clusters appear becomes computationally prohibitive. One way to address this problem is to numerically solve the evolution of different classes of clusters with dedicated methods \cite{jourdan2014efficient, surh2004master}. We present here a generic algorithm that allows a seamless coupling between such methods.\\

\begin{figure}[htb]
	\centering
	\includegraphics[width=\textwidth]{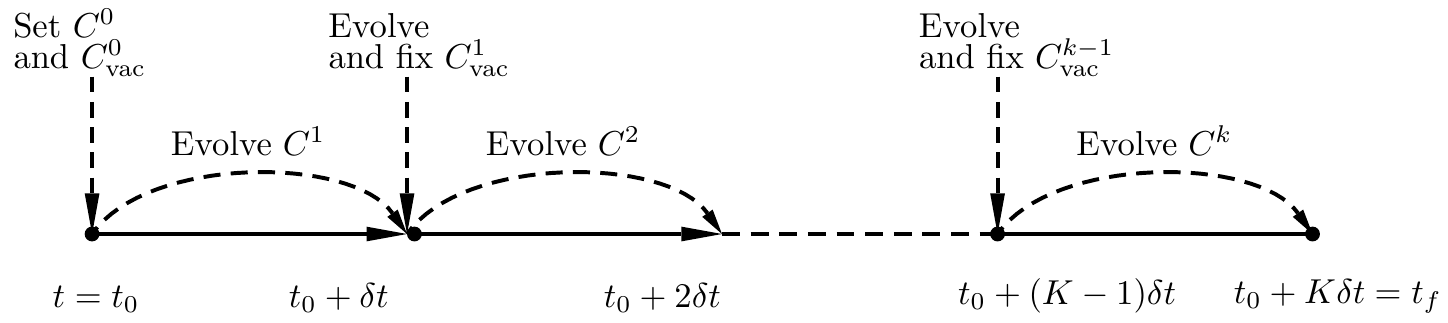}
	\caption{Illustration of the first main idea of the algorithm: splitting between the dynamics of $C_\vac$ and $(C_n)_{n\ge 2}$.}
	\label{fig:schemaP1}
\end{figure}

We propose to first split the dynamics into two elementary dynamics, namely the dynamics of the vacancy concentration $C_\vac$ at fixed concentrations $(C_n)_{n\ge 2}$ (see~\eqref{eq:EPv}) and the dynamics of the cluster concentrations $C = (C_n)_{n\ge2}$ at fixed vacancy concentration $C_\vac$ (see~\eqref{eq:EPn}).  This splitting may be performed after a time $t_0$ corresponding to some initial transient regime where the full set of ODEs~\eqref{eq:EPn}--\eqref{eq:EPv} is integrated by a standard numerical scheme. Let $\delta t$ be a time step and $t_k = t_0 + k \delta t$ for $k\in\{1,\cdots,K\}$ the time at which approximations of the solution are sought, $C_\vac^{k}$ and $C^{k}$ being respectively approximate solutions of $C_\vac(t_k)$ and $C(t_k)$. A good approximation of the cluster dynamics is provided by the following procedure:
\begin{equation}
\left\{
\begin{aligned}
& C^{k} = \mathcal{G}_{\delta t}\left(C^{k-1};C^{k-1}_\vac\right), \\[0.2cm]
& C_\vac^{k} = \mathcal{F}_{\delta t}\left(C_\vac^{k-1}; C^{k}\right),\\
\end{aligned}
\right.
\label{P1}
\tag{P1}
\end{equation}
where $\mathcal{F}_{\delta t}$ and $\mathcal{G}_{\delta t}$ respectively approximate the evolution of~\eqref{eq:EPv} and~\eqref{eq:EPn} over a time step $\delta t$. Figure~\ref{fig:schemaP1} illustrates such a splitting. 

Notice that in the limit $\delta t \to 0$, the problem \eqref{P1} becomes equivalent to the full cluster dynamics~\eqref{eq:EPn}--\eqref{eq:EPv}. The error is determined by the time step $\delta t$ and the quality of the approximations $\mathcal{F}_{\delta t}$ and $\mathcal{G}_{\delta t}$. We quantify the error with respect to $\delta t$ in Section~\ref{sec:resQuasi}, where we also discuss the range of admissible $\delta t$.

\subsubsection{Integrating the vacancy subdynamics}

The numerical scheme $\mathcal{F}_{\delta t}$ in~\eqref{P1} is obtained by approximating the solution of~\eqref{eq:EPv} with $C = (C_n)_{n\ge 2}$ fixed. Let us now discuss how to obtain $C_\vac^k$ from the knowledge of $C_\vac^{k-1}$ and $C^k = (C^k_n)_{n\ge 2}$. Introducing
\begin{equation}
\mathcal{A}^k = \sum_{n\ge 2} \alpha_n C_n^k + \alpha_2 C_2^k,
\label{eq:A}
\end{equation} 
and \begin{equation}
\mathcal{B}^k =  \sum_{n\ge 2} \beta_n C_n^k,
\label{eq:B}
\end{equation}
$C_\vac^k$ is an approximation of the solution of the following dynamics at time $\delta t$:
\begin{equation}
\frac{d{C}_\vac}{dt} = -\beta_1 {C}_\vac^2(t) - \mathcal{B}^k {C}_\vac(t) + \mathcal{A}^k, \quad {C}_\vac(0) = C_\vac^{k-1}.
\label{eq:EPvQuasi}
\end{equation}
The actual numerical method $\mathcal{F}_{\delta t}$ depends on the numerical scheme used to integrate~\eqref{eq:EPvQuasi} (see Section~\ref{sec:Cvac}).

\subsubsection{Integrating the cluster subdynamics}

The numerical scheme $\mathcal{G}_{\delta t}$ in~\eqref{P1} is obtained by approximating the solution of~\eqref{eq:EPn} with $C_\vac$ fixed. Let us now discuss how to obtain $C^k$ from the knowledge of $C_\vac^{k-1}$ and $C^{k-1}$. First notice that, when the vacancy concentration is fixed, the set of rate equations~\eqref{eq:EPn} forms a linear problem, that can be express in matrix form. Denote by $(e_n)_{n \ge 0}$ the basis with components $(e_n)_i = \delta_n(i)$, where $\delta$ is the Kronecker delta. Let $A_0$ be the tridiagonal operator such that:
\begin{equation}
\begin{aligned}
&A_0(C_\vac)e_2 = - (\beta_2 C_\vac + \alpha_2)e_2 + \beta_2 C_\vac e_3, \\[0.2cm]
&A_0(C_\vac)e_n = \alpha_{n}e_{n-1} - (\beta_n C_\vac + \alpha_n)e_n + \beta_{n} C_\vac e_{n+1}.
\end{aligned}
\end{equation}
The operator $A_0$ can also be represented as the following infinite matrix:
%\begin{strip}
\begin{equation}
A_0(C_\vac) = \begin{pmatrix}
-(\beta_2C_\vac+\alpha_2) & \alpha_3 & 0 & 0 & \cdots \\[0.3cm] 
\beta_2C_\vac & -(\beta_3C_\vac+\alpha_3) & \alpha_4 & 0 & \cdots \\[0.3cm] 
0 & \beta_3C_\vac & -(\beta_4C_\vac+\alpha_4) & \alpha_5 & \cdots \\[0.3cm] 
0 & 0 & \beta_4C_\vac & -(\beta_5C_\vac+\alpha_5) & \ddots \\[0.3cm] 
\vdots & \vdots & \vdots & \ddots & \ddots &
\end{pmatrix} 
\end{equation}
%\end{strip}
The approximation $C^k$ is obtained by numerically integrating the following dynamics:
\begin{equation}
\left\{
\begin{aligned}
& \frac{dC}{dt} = A_0(C^{k-1}_\vac)C, \\[0.2cm]
& C(0) = C^{k-1},
\end{aligned}
\right.
\label{P2}
\tag{P2}
\end{equation}
depending on the numerical scheme used to integrate~\eqref{P2}.\\

As previously noticed, a key feature of~\eqref{eq:EPn} is that the dynamics is linear. For any initial condition $C^0$, it is then possible to split the evolution problem~\eqref{P2} into independent evolutions, corresponding to a decomposition of the initial condition $C^0$. The solution is then obtained by summing the independent sub-solutions. If one writes $C^0 = C^{0,a} + C^{0,b}$, then the solution is $C(t) = C^a(t)+C^b(t)$ with $C^z(t)$ the solution of~\eqref{P2} with initial condition $C^{0,z}$ for $z = a,b$. 

\subsubsection{Splitting and decomposition of the dynamics}

\begin{figure*}[htb]
    \centering
    \begin{subfigure}[b]{0.5\textwidth}
        \centering
        \includegraphics[width=\textwidth]{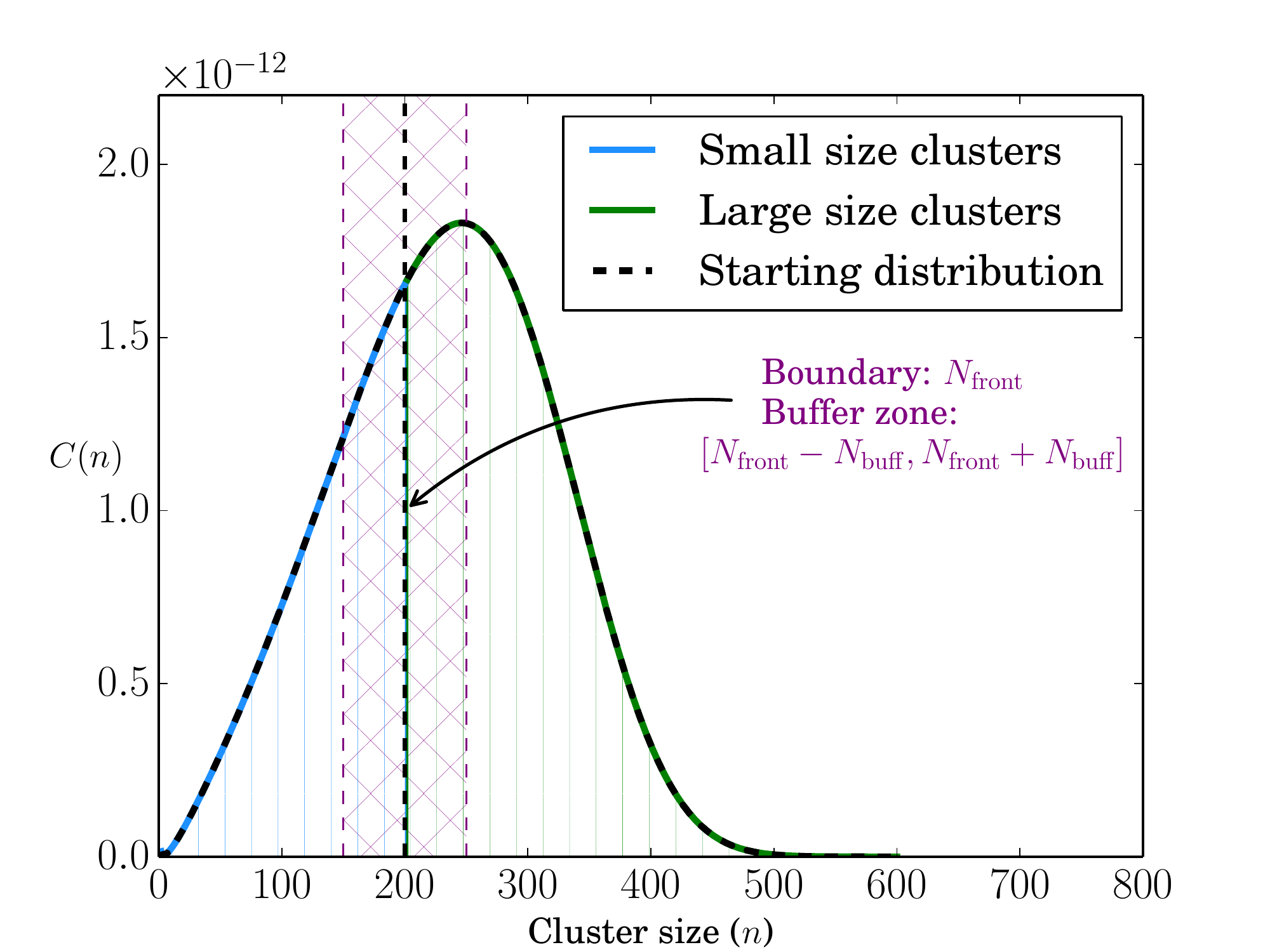}
        \caption{Initial distributions}
    \end{subfigure}%
    ~ 
    \begin{subfigure}[b]{0.5\textwidth}
        \centering
        \includegraphics[width=\textwidth]{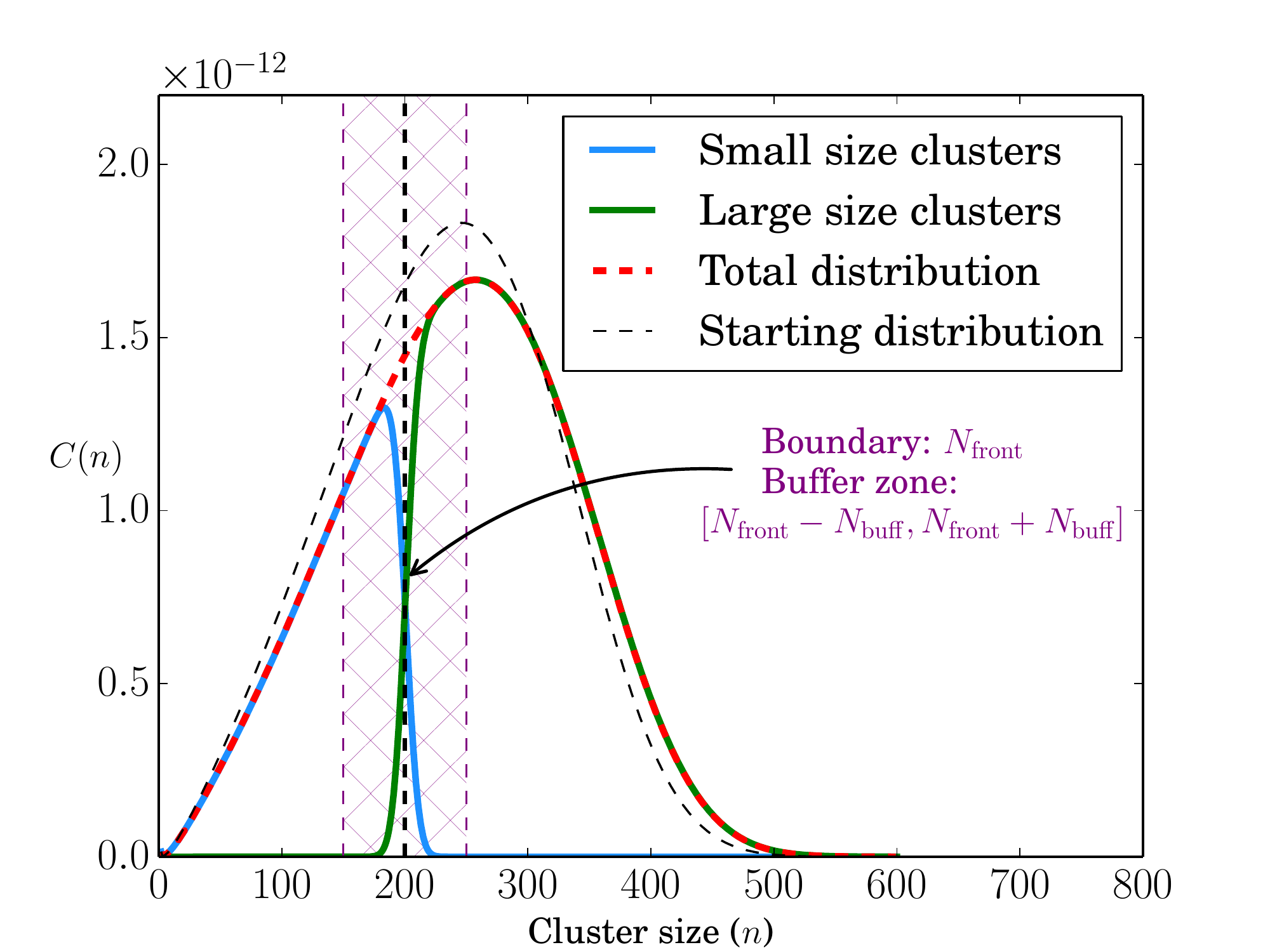}
        \caption{Propagated distributions}
    \end{subfigure}
    \caption{Illustration of the second main idea of the algorithm: the initial distribution is divided into two distributions, which are independently propagated. The buffer zone allows the distributions to overlap and limits the calculation cost since both distributions are propagated on a limited space.}
    \label{fig:Demo}
\end{figure*}

Using the linearity of~\eqref{P2} we choose to separate the evolution of small and large size clusters. With initial conditions $C_\mathrm{small}^{k-1}$ and $C_\mathrm{large}^{k-1}$ such that $C_\mathrm{small}^{k-1} + C_\mathrm{large}^{k-1} = C^{k-1}$, the main problem~\eqref{P1} now writes:
\begin{equation}
\left\{
\begin{aligned}
& C_\mathrm{small}^k = \mathcal{G}^\mathrm{small}_{\delta t}\left(C_\mathrm{small}^{k-1};C_\vac^{k-1}\right), \\[0.2cm]
& C_\mathrm{large}^k = \mathcal{G}^\mathrm{large}_{\delta t}\left(C_\mathrm{large}^{k-1};C_\vac^{k-1}\right), \\[0.2cm]
& C_\vac^{k} = \mathcal{F}_{\delta t}\left(C_\vac^{k-1};C_\mathrm{small}^{k}+C_\mathrm{large}^{k}\right),\\[0.2cm]
\end{aligned}
\right.
\label{P3}
\tag{P3}
\end{equation}
Note that such a decomposition between small and large size clusters allows us to solve the corresponding dynamics with a different numerical scheme (as emphasized by the notations $\mathcal{G}^\mathrm{small}$ and $\mathcal{G}^\mathrm{large}$). It is straightforward and computationally effective to numerically solve rate equations for small size clusters (since they consist of a small number of ODEs), so that many options are available for $\mathcal{G}^\mathrm{small}$. On the other hand, the treatment of large size clusters requires dedicated techniques. We present in Section~\ref{sec:large} two stochastic methods that are highly parallelizable and more appropriate for large size clusters. Figure~\ref{fig:Demo} illustrates the decomposition between small and large size clusters on a single time interval. In Figure~\ref{fig:Demo}.a, the initial distribution is divided into two distributions, one for small size clusters, the other for large size ones. Both distributions are then propagated independently over time and the sum of both propagated distributions (Figure~\ref{fig:Demo}.b) gives us an approximation of the total distribution.

\subsection{Main algorithm}
\label{sec:algo}

From the discussion of Section~\ref{sec:maths}, the introduction of the coupling algorithm is rather straightforward. Let us introduce a final time $t_f$ for the calculation, a frontier $N_\mathrm{front}$ and a buffer zone of size $N_\mathrm{buff}$ for the separation and overlapping of small and large size clusters. The size of the buffer zone is chosen sufficiently small to limit the computational cost, as it allows in particular to reduce the number of ODE to solve. Therefore one has to choose $N_\mathrm{buff}$ and $\delta t$ such that when one propagates the distribution of small size clusters with $C_n = 0$ for $n \ge N_\mathrm{front}$ on a time step $\delta t$, it remains negligible for $n\ge N_\mathrm{front}+N_\mathrm{buff}$. Actually, the distribution around $n$ approximately propagates at an average speed of $\beta_n C_\vac - \alpha_n$.  This property can be observed easily on the Fokker-Planck equation~\eqref{eq:FP}, presented in Section~\ref{sec:FP}, with $F$ acting as the drift term. Therefore $N_\mathrm{buff}$ can be chosen to be of order $(\beta_{N_\mathrm{front}} C_\vac - \alpha_{N_\mathrm{front}})\delta t$. We also set a maximum cluster size to $N_\mathrm{max}$. \\

\begin{figure*}[htb]
	\centering
	\includegraphics[width=\textwidth]{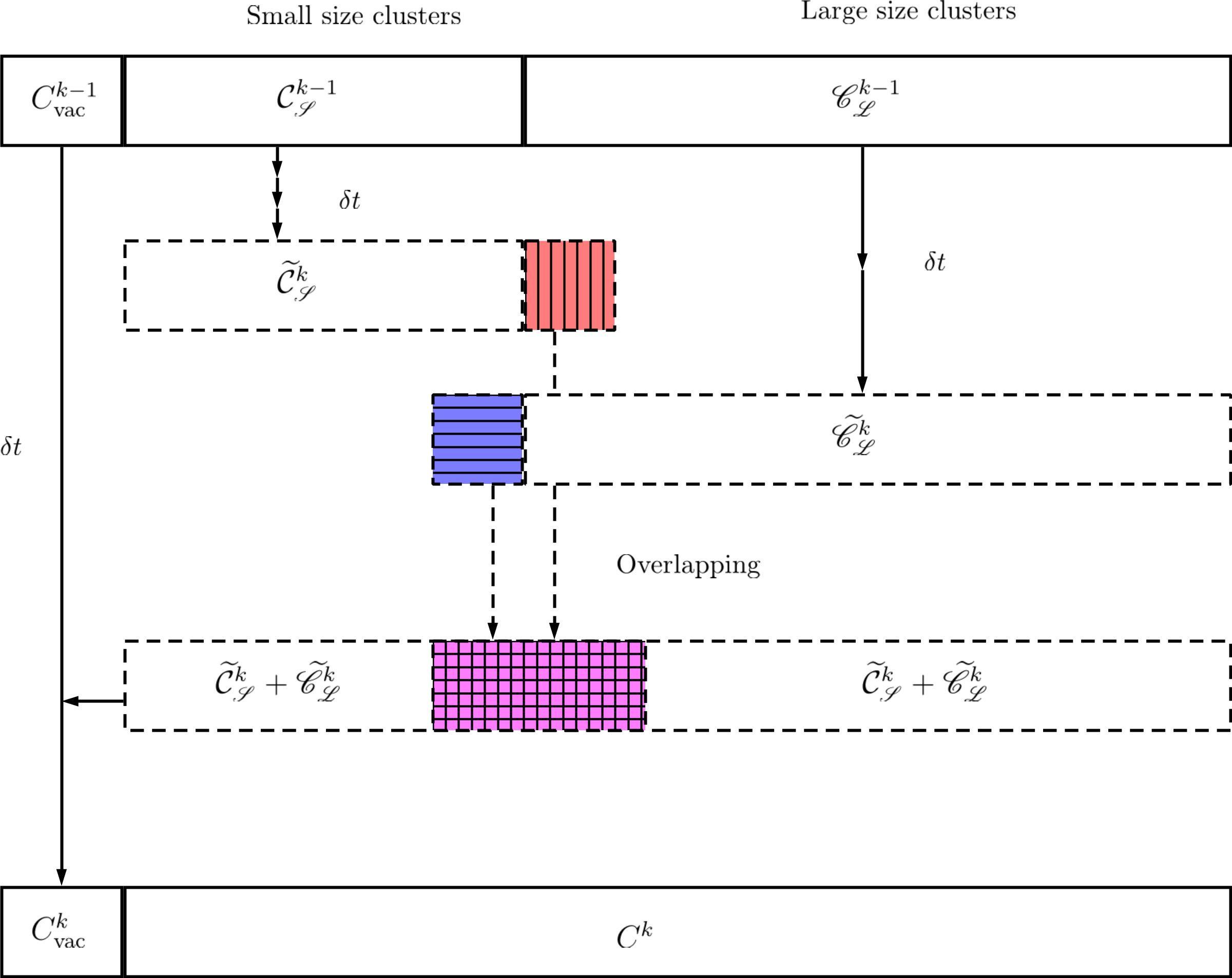}
	\caption{Summary of the algorithm presented in Section~\ref{sec:algo}.}
	\label{fig:schema}
\end{figure*}

As explained in Section~\ref{sec:maths}, we intend to solve with different methods the evolution of small and large size clusters. While the scheme $\mathcal{G}_{\delta t}^\mathrm{small}$ can be as simple as a Euler scheme for ODEs, the scheme $\mathcal{G}_{\delta t}^\mathrm{large}$ is made precise in Section~\ref{sec:large}. The diagram presented in Figure~\ref{fig:schema} summarizes the algorithm presented hereafter.\\

Let $C^0 = (C^0_n)_{n\ge 2}$ be the initial distribution of the cluster concentrations, $\mathcal{C}^0_\mathscr{S} = (C^0_2, \cdots, C^0_\mathrm{N_\mathrm{front}-1},0,\cdots)$ the initial distribution for small size clusters and $\mathscr{C}^0_\mathscr{L} = (0, \cdots, 0, C^0_{N_\mathrm{front}}, C^0_{N_\mathrm{front}+1},\cdots)$ the initial distribution for large size clusters and $C_\vac^0$ the initial vacancy concentration. To compute the solution from a time $k\delta t$ to a time $(k+1)\delta t$, the general algorithm reads as follows:

\begin{enumerate}[(1)]
\item[(0)] Decompose the total distribution between small size and large size clusters:
\begin{equation}
\begin{aligned}   
    &\mathcal{C}^{k}_\mathscr{S} = \left(\widetilde{\mathcal{C}}^{k}_\mathscr{S}(2),\cdots, \widetilde{\mathcal{C}}^{k}_\mathscr{S}(N_\mathrm{front}-1),0,\cdots\right),\\[0.2cm]
    &\mathscr{C}^{k}_\mathscr{L} = \left(0,\cdots,0,\widetilde{\mathscr{C}}^{k}_\mathscr{L}(N_\mathrm{front}), \widetilde{\mathscr{C}}^{k}_\mathscr{L}(N_\mathrm{front}+1),\cdots\right).\\[0.2cm]
\end{aligned}
\label{algo:M0}
\tag{M0}
\end{equation}

\item Compute $\widetilde{\mathcal{C}}^{k+1}_\mathscr{S}$ on $\{2,\cdots,N_\mathrm{front}+N_\mathrm{buff}\}$ by integrating the ODE~\eqref{eq:EPn} with initial condition $\mathcal{C}^k_\mathscr{S}$ equal to 0 for $n\ge N_\mathrm{front}$:\\
\begin{equation}
    \widetilde{\mathcal{C}}^{k+1}_\mathscr{S} = \mathcal{G}_{\delta t}^\mathrm{small}\left(\mathcal{C}^{k}_\mathscr{S};C_\vac^k\right).
    \label{algo:M1}
\tag{M1}
\end{equation}

\item Compute $\widetilde{\mathscr{C}}^{k+1}_\mathscr{L}$ on $\{N_\mathrm{front}-N_\mathrm{buff},\cdots,N_\mathrm{max}\}$ by a (possibly approximate) dynamics for large size clusters, with initial condition $\mathscr{C}^{k}_\mathscr{L}$ equal to 0 for $n\le N_\mathrm{front}-1$:\\
\begin{equation}
    \widetilde{\mathscr{C}}^{k+1}_\mathscr{L} = \mathcal{G}_{\delta t}^\mathrm{large}\left(\mathscr{C}^{k}_\mathscr{L};C_\vac^k\right),
    \label{algo:M2}
\tag{M2}
\end{equation}

\item Compute the total distribution $C^{k+1}$:
\begin{equation}
C^{k+1} = \widetilde{\mathcal{C}}^{k+1}_\mathscr{S} + \widetilde{\mathscr{C}}^{k+1}_\mathscr{L},
\label{algo:M3}
\tag{M3}
\end{equation}

\item Update the vacancy concentration $C_\vac^{k+1}$:
\begin{equation}
    C_\vac^{k+1} = \mathcal{F}_{\delta t}\left(C_\vac^k;C^{k+1}\right).
\label{algo:M4}
\tag{M4}
\end{equation}
\end{enumerate}
\noindent Let us make some general remarks:
\begin{itemize}
\item[-] Before the distribution reaches the border $N_\mathrm{front}$, equations~\eqref{eq:EPn} and~\eqref{eq:EPv} can be solved by an ODE scheme without any splitting between $C_\vac$ and $(C_n)_{n\ge 2}$. This may be important for the initial transient regime where $C_\vac$ rapidly evolves.
%\item[-] The accuracy of such an algorithm depends of the first splitting between $C_\vac$ and $(C_n)_{n\ge 2}$. In particular, a slow evolution of $C_\vac$ will allow to take large time steps $\delta t$. This point will be discussed in Section~\ref{sec:resQuasi}. 
\item[-] An alternative way to handle the boundary and the buffer zone is to introduce adaptative time steps $\delta t$ for a fixed value $N_\mathrm{buff}$. These steps would be limited by the requirement that $\mathcal{C}_\mathscr{S}(N_\mathrm{front}+N_\mathrm{buff})$ should not grow to a value that is not negligible any more.
\item[-] Since Steps~\eqref{algo:M1} and~\eqref{algo:M2} are independent, it is possible to switch those steps or to perform them simultaneously.
\end{itemize}

\section{Discretization of the large size cluster subproblems}
\label{sec:large}

As we want to avoid the computation of a large number of ODEs, we present two methods that allow for a stochastic and parallelizable solving of the evolution of large size clusters (Step~\eqref{algo:M2} in the algorithm of Section~\ref{sec:algo}). The first method, called Birth-Death process approach, relies on an exact formulation of the set of rate equations when $C_\vac$ is fixed, but is better suited for clusters of intermediary sizes. The second method, called Fokker-Planck approach, is based on an approximation by a partial differential equation of the set of rate equations for large size clusters, which is more appropriate for sufficiently large clusters.

\subsection{The Birth-Death process approach}
\label{sec:BD}

\subsubsection{Model presentation}

The first stochastic approach we consider in this work is a Birth-Death process approach. Assuming that the vacancy concentration $C_\vac$ is fixed, the rate equations~\eqref{eq:EPn} are equivalent to a forward-Kolmogorov equation of a Markov process. Such an observation was already made by Goodrich \cite{goodrich1964nucleation}. Indeed a cluster of size $n$ either emits a vacancy and reduces to a cluster of size $n-1$ or absorbs a vacancy and grows to a cluster of size $n+1$. Such a behaviour can be seen as the death (or disappearing) of a vacancy amidst a population of vacancies of size $n$, or the birth of a vacancy amidst this population. Since the rate of absorption depends on $C_\vac$, the Markov process is actually a time-dependent birth-death process. It is described as follow: consider a population of size $N(t)$ at a time $t$ and a small time step $\delta \tau$. The transition probabilities for $n \ge 2$ are given by
\begin{equation}
\begin{aligned}
    &\mathbb{P}\left[N(t+\delta \tau) = n+1 | N(t) = n\right] = \beta_n C_\vac \delta \tau + \mathrm{O}(\delta \tau),\\[0.2cm]
    &\mathbb{P}\left[N(t+\delta \tau) = n-1 | N(t) = n\right] = \alpha_n \delta \tau + \mathrm{O}(\delta \tau),\\[0.2cm]
    &\mathbb{P}\left[N(t+\delta \tau) = n | N(t) = n\right] = 1 - (\beta_n C_\vac +\alpha_n) \delta \tau + \mathrm{O}(\delta \tau).
\end{aligned}
\label{eq:jump}
\end{equation}
Let $p(t,n) = \mathbb{P}(N(t) = n)$ be the probability to be in state $n$ at a time $t$. The evolution of $(p(t,n))_{n\ge 2}$ is then governed by the following dynamics, the forward-Kolmogorov equation:
\begin{equation}
    \frac{dp}{dt}(t,n) = \beta_{n-1} p(t,n-1) C_\vac - \left( \beta_n C_\vac + \alpha_n \right) p(t,n) + \alpha_{n+1} p(t,n+1),
    \label{eq:BirthDeath}
\end{equation}
with initial conditions
\begin{equation}
p(0,n) = 1, \quad \text{and}\quad  p(0,i) = 0, \quad i\neq n,
\end{equation}
for a population initially of size $n$. There exist many algorithms that allow to compute an approximation of $p$ at each time, assuming the concentration $C_\vac$ is known. In the sequel, we introduce a simple but rather efficient one, where we propagate particles according to the law of the jump process~\eqref{eq:jump}. Each particle represents a cluster of size $n$ and evolves independently on a time step $\delta t$ (on which $C_\vac$ is constant), similarly to a population that experiences birth and death of its elements. Due to the independence of each particle, the method is highly parallelizable. Moreover, as we only consider large size clusters, we do not suffer from high frequency events due to the small clusters behaviour as illustrated in Figure~\ref{fig:nu}, where the characteristic jump time is really small for small size clusters.

\subsubsection{Birth-Death algorithm: Step~\eqref{algo:M2}}
\label{sec:BDalgo}

The Birth-Death process approach interprets the set of rate equations as a set of forward-Kolmogorov equations and its solution is therefore a probability distribution, denoted by $(p(t,n))_{n\ge 2}$ such that $p(t,n) \ge 0$ and $\sum_{n=2}^{\infty} p(t,n) = 1$. The total concentration \begin{equation}
M_\mathrm{tot} = \sum_{n=2}^{\infty} \mathscr{C}_\mathscr{L}(n)
\label{eq:Mtot}
\end{equation}
should be stored in order to rescale the probability $p$ and get the concentration as $\mathscr{C}_\mathscr{L} = M_\mathrm{tot} p$.\\

In order to compute the law $p(t,n)$, starting from a distribution $p_0$, the method we propose generates a large number $N_\mathrm{sim}$ of particles $(X_n)_{1\le n\le N_\mathrm{sim}}$ sampled according to $p_0$. There exist various methods to sample from a multinomial distribution ($p_0$ is discrete), see for instance~\cite{Malefaki2007}. These particles are then propagated according to the jump process associated with the transition rates~\eqref{eq:jump}. For a particle in state $n$, its jump frequency is given by 
\begin{equation}
\nu(n) = \beta_n C_\vac + \alpha_n
\label{eq:nu}
\end{equation} 
(\textit{i.e.} the time of the next jump follows an exponential distribution $\mathcal{E}$ of rate $\nu(n)$) and, when it jumps, the particle reaches either the state $n-1$ with probability $\alpha_n/(\beta_n C_\vac + \alpha_n)$ or the state $n+1$ with probability $\beta_n C_\vac/(\beta_n C_\vac + \alpha_n)$. We denote by $\xi(x,\tau,u)$ the function which gives the new state as a function of the previous one $x$ and the two random numbers used in the procedure. Here,  $\tau$ is a random time sampled from an exponential distribution of parameter $\nu(x)$ and $u$ is a random number sampled from a uniform distribution on $[0,1]$ allowing us to choose between the state $n-1$ and $n+1$.\\

There is in fact no notion of time step in this algorithm, but rather a final time $\delta t$. The algorithm summarized as $\mathcal{G}_{\delta t}^\mathrm{large}(\mathscr{C}^k_\mathscr{L};C_\vac^k)$ in equation~\eqref{algo:M2} at a step $k$ reads as follows:

\begin{enumerate}[(1)]
\item Sample $N_\mathrm{sim}$ particles according to the initial distribution $p_0^k(n) = M_\mathrm{tot}^{-1} \mathscr{C}_\mathscr{L}^k(n)$ as:
\begin{equation}
(x_1^0,\cdots,x_{N\mathrm{sim}}^0) \sim p_0^k;
\label{algo:B1}
\tag	{B1}
\end{equation}

\item Propagate the particles until $(k+1)\delta t$:
\begin{enumerate}[(a)]
\item Associate the $\ell$-th particle (denoted by $x_\ell$) with a time $\tau_\ell^{k}$ that is initially set to $k\delta t$;

\item Propagate independently all particles:
\begin{enumerate}[(i)]
\item first sample a jump time:
\begin{equation}
\begin{aligned}
& \text{Compute the jump frequency :\ } \nu(x_\ell) = \beta_{x_\ell}C_\vac^{k} + \alpha_{x_\ell}; \\[0.2cm]
& \text{Sample a jump time\ } \delta\tau_\ell \sim \mathcal{E}(\nu(x_{\ell})); \\[0.2cm]
& \text{Update time\ }\tau_\ell^{k} \longleftarrow \tau_\ell^{k} + \delta \tau_\ell;\\[0.2cm]
\end{aligned}
\label{algo:B2.a}
\tag{B2.a}
\end{equation}

\item and then propagate until $(k+1)\delta t$:
\begin{equation}
\begin{aligned}
& \text{While\ }\tau_\ell^k \le (k+1)\delta t,\ \text{do:}\\[0.2cm]
& \text{\ - Sample:\ } u \sim U; \\[0.2cm]
& \text{\ - Propagate:\ } x_\ell \longleftarrow \xi(x_\ell,\delta\tau_\ell,u); \\[0.2cm]
& \text{\ - Sample the next jump time\ } \delta\tau_\ell \sim \mathcal{E}(\nu(x_\ell)); \\[0.2cm]
& \text{\ - Update the time as\ }\tau_\ell^{k} \longleftarrow \tau_\ell^{k} + \delta \tau_\ell;
\end{aligned}
\label{algo:B2.b}
\tag{B2.b}
\end{equation}
\end{enumerate}
\end{enumerate}

\item Compute the concentration $\widetilde{\mathscr{C}}^{k+1}_\mathscr{L}$ at time $(k+1)\delta t$:
\begin{equation}
\widetilde{\mathscr{C}}^{k+1}_\mathscr{L}(n) = \frac{M_\mathrm{tot}}{N_\mathrm{sim}}\sum_{\ell=1}^{N_\mathrm{sim}} \mathds{1}_{n}(x_{\ell}),
\label{algo:B3}
\tag{B3}
\end{equation}
for $n = N_\mathrm{front}-N_\mathrm{buff},\cdots,N_\mathrm{max}$.
\end{enumerate}

Notice that at a step $k > 0$ of the main algorithm, the Step~\eqref{algo:B1} can be performed without actually resampling $N_\mathrm{sim}$ particles. To this end, from the previous step $k-1$, one just needs to add the particles coming from the part of the distribution of small size clusters that spills out in the large size clusters zone and delete the ones that are added to the small size cluster distribution. In order to keep the number of particles constant, we either randomly suppress some of the particles if there are more than $N_\mathrm{sim}$ particles, or duplicate some of them if there are less than $N_\mathrm{sim}$ particles.\\

Such an algorithm differs from the standard SSA procedure in that all particles are handled independently (which itself comes from the fact that the forward-Kolmogorov equation on $p$ is linear). Parallelizing the scheme is straightforward in the present situation and results in a significant improvement in term of wall-clock time. Moreover this method becomes exact in the limit $N_\mathrm{sim} \to +\infty$. Nevertheless the frequency $\nu(n)$ increases with $n$ and this might reduce the efficiency of the method for very large clusters. In these situations, Fokker-Planck methods should be used instead.

\subsection{The Fokker-Planck approach}
\label{sec:FP}

\subsubsection{Model presentation}

Assuming that the size $n$ of the cluster is large enough, it has been noticed, assuming that the concentration $C_\vac$ is known at each time, that rate equations can be approximated with a good approximation by a single Fokker-Planck equation~\cite{goodrich1964nucleation,wolfer1977theory}:
\begin{equation}
    \frac{\partial \mathscr{C}}{\partial t} = - \frac{\partial (F \mathscr{C})}{\partial x} + \frac{1}{2} \frac{\partial^2 (D\mathscr{C})}{\partial x^2},
    \label{eq:FP}
\end{equation}
where $F(t,x) = \beta(x)C_\vac(t) - \alpha(x)$ and $D(t,x) = \beta(x)C_\vac(t) + \alpha(x)$. The scalar field $\mathscr{C}$ acts as a concentration which is continuous in space, with:
\begin{equation}
    C_n(t) \simeq \mathscr{C}(t,n)  \quad \text{for}\quad  n \gg 1.
\end{equation}
The size $x$ of the cluster now plays the role of a spatial coordinate. When only one type of defect is considered, such a partial differential equation (PDE) is one-dimensional in space. In this situation, there exist good solvers to efficiently simulate such equations on large scales problems. Jourdan \textit{et al.} \cite{jourdan2016accurate} have proposed an efficient method (based on a finite volume formulation) to numerically solve the Fokker-Planck equation when it is coupled to rate equations.\\

Nevertheless, when two or more types of defects are introduced, \textit{i.e.} when a cluster is identified by a $m$-tuple $(n_1,n_2,\cdots,n_m)$, the Fokker-Planck equation is $m$-dimensional. It then becomes computationally prohibitive to solve it with deterministic mesh-based methods due to the curse of dimensionality (the number of discretization unknowns grow exponentially with the dimension). Stochastic methods are much more appropriate in such situations. The Fokker-Planck equation is related to a stochastic differential equation, called Langevin dynamics. Let $(X_t)_{t\ge 0}$ be the stochastic process  
\begin{equation}
    dX_t = F(t,X_t)dt + \sqrt{D(t,X_t)} dW_t,
    \label{eq:Langevin}
\end{equation}
where $W_t$ is a standard Wiener process. Then the law $p(t,x)$ of $X_t$ satisfies the Fokker-Planck equation~\eqref{eq:FP}. Therefore, by simulating a large number of trajectories for the process $X_t$, one can obtain a good approximation of the law $p$, \textit{i.e.} the solution of the Fokker-Planck equation~\eqref{eq:FP}. Since the trajectories are independent of each other, this stochastic method is also highly parallelizable.
%This approach exploits the connection between the Fokker-Planck equation and a %stochastic differential equation.

\subsubsection{Fokker-Planck algorithm: Step~\eqref{algo:M2}}

It is important to note that using the stochastic representation of the Fokker-Planck equation~\eqref{eq:FP} allows to evolve a probability $p(t,x)$ such that $\int p(t,x)dx = 1$. As in the Birth-Death process approach, the concentration $M_\mathscr{L}$ of large size clusters~\eqref{eq:Mtot} should therefore be stored
in order to rescale the probability $p$ and obtain the concentration $\mathscr{C}_\mathscr{L}$. \\

Let us introduce an interpolation operator $\mathcal{I}$ that transforms a discrete distribution into a continuous one (such as a linear interpolation of the values at integers). The given initial condition $\mathscr{C}^0_\mathscr{L}$ is associated with an initial density 
\begin{equation}
p_0(x) = \frac{\mathcal{I}(\mathscr{C}^0_\mathscr{L})(x)}{\int_{N_\mathrm{front}}^{\infty} \mathcal{I}(\mathscr{C}^0_\mathscr{L})(y)dy},
\end{equation}
and with a fixed $C_\vac$, we formulate the problem as
\begin{equation}
\left\{
\begin{aligned}
&\text{Find the law } p(t,x) \text{ of } (X_t)_{t\ge0} \text{ solution of}\\[0.2cm]
&dX_t = F(X_t)dt + \sqrt{D(X_t)}dW_t,\\[0.2cm]
&p(0,x) = p_0(x),
\end{aligned}
\right.
\end{equation}
where $F, D$ are defined after~\eqref{eq:FP}. In order to approximate the law $p(t,x)$ one can generate a large number of trajectories for $X$ and use various methods such as histograms or kernel density estimators to construct an empirical density. Let us therefore introduce the number $N_\mathrm{sim}$ of Langevin trajectories, $\chi$ a kernel function and $h$ a smoothing parameter. We will in particular need to sample the initial distribution $p_0$ with $N_\mathrm{sim}$ Langevin particles. Here we use a Metropolis algorithm~\cite{metropolis1953, robert2013monte}, starting from $N_\mathrm{front}+N_\mathrm{buff}$, with a uniform proposal distribution of support $[-\alpha,\alpha]$, with $\alpha$ chosen such that the acceptance ratio is around $0.5$. Using the kernel density approach, the law of $X_t$ is approximated by
\begin{equation}
p(t,x) = \frac{1}{N_\mathrm{sim}h}\sum_{\ell=1}^{N_\mathrm{sim}} \chi\left(\frac{X_t^\ell - x}{h}\right),
\end{equation}
where $(X_t^\ell)_{1\le \ell\le N_\mathrm{sim}}$ are trajectories of the process $X$. Finally to propagate the Langevin dynamics, we use a numerical scheme $\psi(x,\Delta t^L,G)$, here a Euler-Maruyama scheme, with a time step $\Delta t^L$: 
\begin{equation}
\psi\left(x,\Delta t^L,G\right) = x + F(x)\Delta t^L + \sqrt{D(x)\Delta t^L}G,
\end{equation} 
where $G$ is a standard Gaussian random variable. The time step $\Delta t^L$ is such that $\delta t = K^L\Delta t^L$ for some $K^L\ge 1$. \\

For the Fokker-Planck approach, the algorithm summarized as $\mathcal{G}_{\delta t}^\mathrm{large}(\mathscr{C}^k_\mathscr{L};C_\vac^k)$ in equation~\eqref{algo:M2} at a step $k$ writes:

\begin{enumerate}[(1)]
\item Sample $N_\mathrm{sim}$ particles according to the initial distribution $p_0^k(x) = \frac{\mathcal{I}(\mathscr{C}^k_\mathscr{L})(x)}{\int_{N_\mathrm{front}}^{\infty} \mathcal{I}(\mathscr{C}^k_\mathscr{L})(y)dy}$ as:
\begin{equation}
\left(x^0_1,\cdots,x^0_{N_\mathrm{sim}}\right) \sim p_0^k;
\label{algo:L1}
\tag{L1}
\end{equation}

\item Propagate in time the Langevin particles for $j = 0,\cdots,K_L-1$:
\begin{equation}
\left(x^{j+1}_1,\cdots,x^{j+1}_{N_\mathrm{sim}}\right) = \left(\psi\left(x^{j}_1,\Delta t^L,G_1^j\right),\cdots,\psi\left(x^{j}_{N_\mathrm{sim}},\Delta t^L,G_{N_\mathrm{sim}}^j\right)\right);
\label{algo:L2}
\tag{L2}
\end{equation}

\item Compute the concentration $\widetilde{\mathscr{C}}^{k+1}_\mathscr{L}$:
\begin{equation}
\widetilde{\mathscr{C}}^{k+1}_\mathscr{L}(n) = \frac{M_\mathrm{tot}}{N_\mathrm{sim}h}\sum_{\ell=1}^{N_\mathrm{sim}} \chi\left(\frac{x_\ell^{N_L}-n}{h}\right),
\label{algo:L3}
\tag{L3}
\end{equation}
for $n = N_\mathrm{front}-N_\mathrm{buff},\cdots,N_\mathrm{max}$.

\end{enumerate}

Once again the stochastic particles are independent and a parallelization of the method is straightforward. Moreover the same remark as in Section~\ref{sec:BDalgo} holds for the Step~\eqref{algo:L1}, to avoid a full resampling and simply update the population size once the distributions have been separated again into distributions of small and large clusters.

\section{Approximating the dynamics of $C_\vac$}
\label{sec:Cvac}

The value of $C_\vac$ needs to be calculated at each multiple of the time increment $\delta t$ in the Step~\eqref{algo:M4} of the main algorithm. This is encoded in the numerical method $C_\vac^{k} = \mathcal{F}(C_\vac^{k-1};C^k)$. We present here three methods to this end.

\subsection{Decomposition into elementary integrable ODEs}
\label{sec:CvacODE}
The first method consists in solving the ODE~\eqref{eq:EPvQuasi} with fixed concentrations $(C_n)_{n\ge 2}$. A direct solving of the ODE with standard schemes is not appropriate in our case because the values $\mathcal{A}^k$ and $\mathcal{B}^k$ are fluctuating due to the stochastic nature of our problem for large size clusters. Since the right hand side term of~\eqref{eq:EPvQuasi} is the difference of two large terms and is observed to be small when integrating the full cluster dynamics (see Figure~\ref{fig:TdCv}), small fluctuations create large instabilities. In order to develop a stable numerical scheme we recommend a decomposition of the evolution into two integrable parts (an affine part and a nonlinear one). For fixed $\mathcal{A}^k$ and $\mathcal{B}^k$ (defined in~\eqref{eq:A} and~\eqref{eq:B}) we split~\eqref{eq:EPvQuasi} into the affine part
\begin{equation}
\frac{dC^\mathrm{L}_\vac}{dt} = - \mathcal{B}^k C^\mathrm{L}_\vac + \mathcal{A}^k,
\end{equation}
and the non-linear one
\begin{equation}
\frac{dC^\mathrm{NL}_\vac}{dt} = - \beta_1 \left(C^\mathrm{NL}_\vac\right)^2.
\end{equation}
Both equations admit analytic solutions, namely:
\begin{equation}
C^\mathrm{L}_\vac(t+t_\mathrm{init}) = \left(C^\mathrm{L}_\vac(t_\mathrm{init})-\frac{\mathcal{A}^k}{\mathcal{B}^k} \right)\exp(-\mathcal{B}^k t) + \frac{\mathcal{A}^k}{\mathcal{B}^k},
\end{equation}
\begin{equation}
C^\mathrm{NL}_\vac(t+t_\mathrm{init}) = \frac{C^\mathrm{NL}_\vac(t_\mathrm{init})}{1+2\beta_1 t C^\mathrm{NL}_\vac(t_\mathrm{init})}.
\end{equation}
To compute the solution one may adopt either a first or second order scheme. To integrate the dynamics with a time step $\Delta t$, such that $\delta t = J\Delta t$, the second order scheme writes
\begin{equation}
\left\{
\begin{aligned}
& C_\vac^{k-1,j+1/2} = \left(C^{k,j}_\vac-\frac{\mathcal{A}^k}{\mathcal{B}^k} \right)\exp\left(-\frac{\mathcal{B}^k \Delta t}{2}\right) + \frac{\mathcal{A}^k}{\mathcal{B}^k}, \\[0.2cm]
& \widetilde{C}_\vac^{k-1,j+1} = \frac{C^{k,j+1/2}_\vac}{1+\beta_1 \Delta t C^{k,j+1/2}_\vac}, \\[0.2cm]
& C_\vac^{k-1,j+1} = \left(\widetilde{C}_\vac^{k,j+1}-\frac{\mathcal{A}^k}{\mathcal{B}^k} \right)\exp\left(-\frac{\mathcal{B}^k \Delta t}{2}\right) + \frac{\mathcal{A}^k}{\mathcal{B}^k}.
\end{aligned}
\right.
\label{eq:Strang}
\end{equation}
The value $C_\vac^{k}$ is set to $C^{k-1,J}_\vac$ in Step~\eqref{algo:M4} of the main algorithm. 

\subsection{Quasi-stationary limit}

The second method consists in making a stronger assumption on the behaviour of $C_\vac$, namely that
\begin{equation}
\frac{dC_\vac}{dt} \simeq 0.
\end{equation}
This situation occurs in many physical systems; see for instance Figure~\ref{fig:TdCv} for an illustration. The vacancy concentration is then given by the positive solution of the second order equation 
\begin{equation}
-2\beta_1 C_\vac^2 - \mathcal{B}^k C_\vac + \mathcal{A}^k = 0.
\end{equation}
The two solutions of this equation are
\begin{equation}
r_{\pm} = -\frac{\mathcal{B}^k}{4\beta_1} \pm \frac{\sqrt{\left(\mathcal{B}^k\right)^2 + 8\beta_1\mathcal{A}^k}}{4\beta_1}.
\label{eq:rpm}
\end{equation}
Since $r_{-} < 0$, the only physical solution is $r_{+} > 0$. The second way of implementing~\eqref{algo:M4} is therefore given by
\begin{equation}
C_\vac^{k+1} = \frac{2\mathcal{A}^k}{\mathcal{B}^k}\left( 1 + \sqrt{1+\frac{8\beta_1\mathcal{A}^k}{\left(\mathcal{B}^k\right)^2}} \right)^{-1},
\label{eq:QuasiStat}
\end{equation}
which is equivalent to $r_+$ in~\eqref{eq:rpm}, although numerically more stable when $\mathcal{A}$ and $\mathcal{B}$ are large.

\subsection{Mass conservation}

The third method is based on the preservation of the total quantity of matter (see~\eqref{eq:Qtot}):
\begin{equation}
C_\vac + \sum_{n\ge 2} nC_n(t) = C_\mathrm{init}.
\label{eq:Mass}
\end{equation} 
The computation of $C_\vac$ given the concentrations $(C_n)_{n\ge2}$ is then straightforward:
\begin{equation}
C^{k}_\vac = C_\mathrm{init} - \sum_{n\ge 2} n C_n^{k}.
\label{eq:Mass}
\end{equation}
This physical property can still be used under irradiation as the conservation law~\eqref{eq:Mass} is modified to take into account an incoming flux: 
\begin{equation}
\sum_{n\ge 1} nC_n(t) = C_\mathrm{init} + \int_0^t \mathscr{P}_\vac(s)ds,
\end{equation}
where $\mathscr{P}_\vac$ is a creation rate of vacancy clusters.

\section{Results}
\label{sec:results}

All the simulations reported below are performed using the parameters of \citep{ovcharenko2003gmic++}, which are summarized in Table~\ref{tab:param}.
\begin{table*}[htb]
\centering
\begin{tabular}{ll}
\hline 
Temperature ($T$) & $823$ K \\ 
Atomic volume ($V_\mathrm{at}$) & $1.205 \times 10^{-29}$ m$^3$ \\ 
Vacancy formation energy ($E_\vac^f$) & $1.7$ eV \\
Vacancy migration energy ($E_\vac^m$) & $1.1$ eV \\
Vacancy diffusion coefficient ($D_\vac$) & $10^{-6} \exp(-E_\vac^m/(k_B T))$ m$^2$s$^{-1}$ \\
Surface energy ($\gamma$) & $1.0$ J/m$^2$ \\
Concentration of quenched-in vacancies ($C_\mathrm{init}$) & $10^{-7}$ atom$^{-1}$ \\
Kernel function ($K$) & $(2\pi)^{-1/2}\exp(-x^2/2)$ \\
Smoothing parameter ($h$) & $0.5$ \\
\hline 
\end{tabular} 
\caption{Physical parameters for a nickel-like metal}
\label{tab:param}
\end{table*}
The final time of computation $t_f$ as well as the time interval $\delta t$ and the time steps $\Delta t^M$, $\Delta t^L$ are specified for each results. Numerical analysis in simple cases show that, besides the error inherent to stochastic methods (related to the variance), there is an additional systematic error, due to the Fokker-Planck approximation, of order $N_\mathrm{front}^{-2/3}$. A more thorough study of approximation errors will be provided in a future paper. The simulations we performed (not reported here) show that, with $N_\mathrm{front} = 200$, the error due to the Fokker-Planck approximation is negligible compared to the statistical error as long as the number of particles $N_\mathrm{sim}$ does not exceed $10^8$. The buffer zone is of size $2N_\mathrm{buffer} = 100$ and therefore extends from $n = 150$ to $n = 250$.

\subsection{On the quasi-stationary assumption}
\label{sec:resQuasi}

We first show that under the assumption that $C_\vac$ reaches a quasi-stationary state, the problem~\eqref{P1} is an excellent approximation of the original cluster dynamics problem~\eqref{eq:EPn}--\eqref{eq:EPv} even for large time intervals $\delta t$. Let us introduce a time-dependent function $\mathcal{T}$ defined by
\begin{equation}
\mathcal{T}(t) = \left| \frac{1}{C_\vac(t)}\frac{dC_\vac}{dt}\right|^{-1}.
\label{eq:defT}
\end{equation} 
The function $\mathcal{T}$ acts as a characteristic time. Since
\begin{equation}
C_\vac(t+\delta t) = C_\vac(t) + \frac{dC_\vac}{dt}(t) \delta t + \mathrm{O}(\delta t^2),
\end{equation}
on a time step $\delta \tau \ll \mathcal{T}$, the variation of $C_\vac$ is relatively small. The condition $\delta t \ll \mathcal{T}$ indicates the relevant orders of magnitude of $\delta t$.\\

We compute a reference solution by solving the full ODE problem~\eqref{eq:EPn}--\eqref{eq:EPv} without any approximation other than the integration scheme. We use a second order Euler-Heun numerical scheme. Starting from the initial condition~\eqref{eq:init}, the system first goes through a nucleation stage (for which we use a small time step $\Delta t^M = 10^{-5}$ s), before it enters a growth regime where we start to observe the quasi-stationary state of $C_\vac$ (characterized by $\frac{dC_\vac}{dt} \simeq 0$). This state is rapidly observed and is reached before clusters grow beyond $N_\mathrm{front}$.

\begin{figure*}[htb]
    \centering
    \begin{subfigure}[b]{0.5\textwidth}
        \centering
        \includegraphics[width=\textwidth]{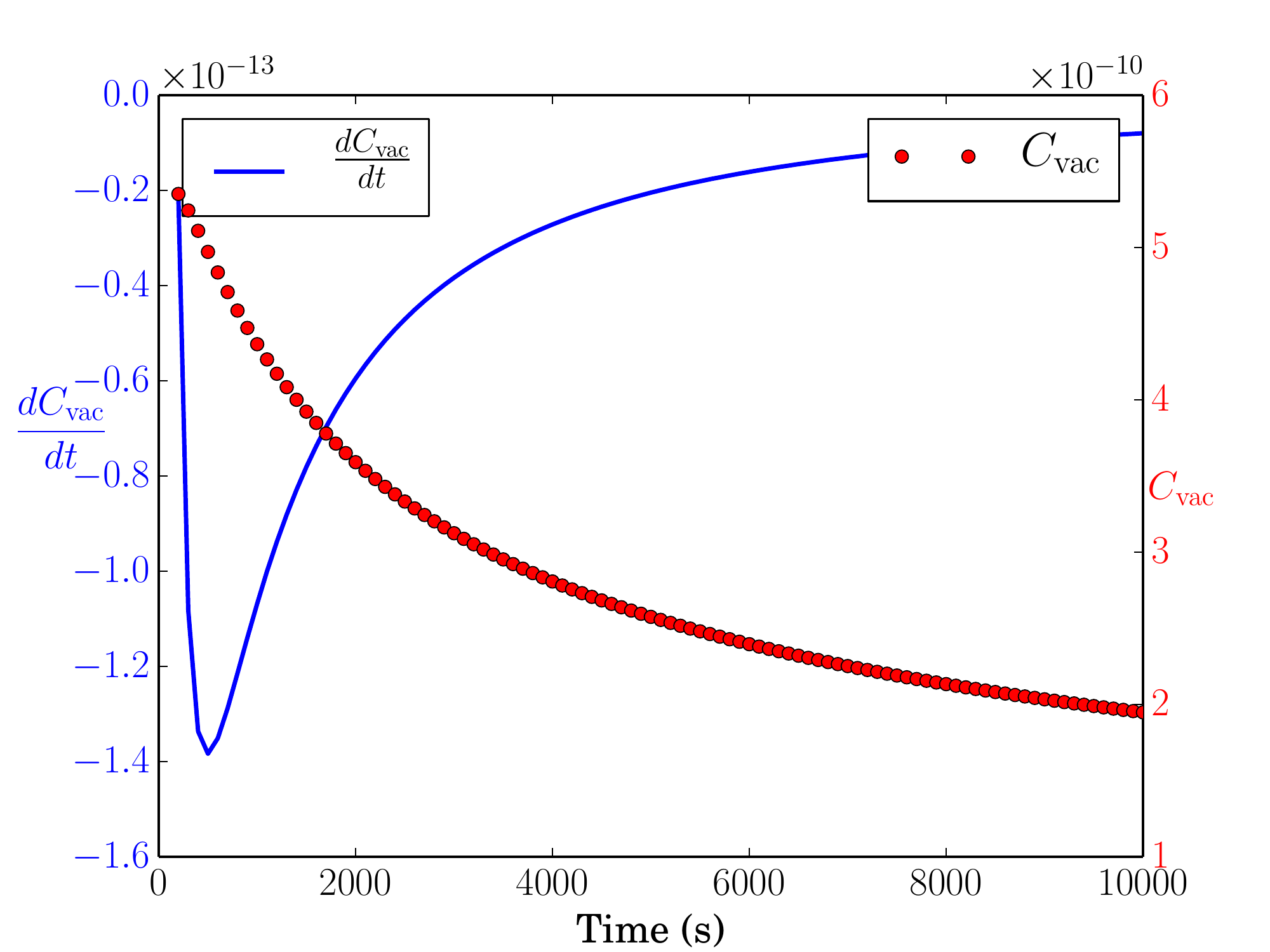}
        %\caption{$C_\vac$ and its derivative as a function of time.}
    \end{subfigure}%
    ~
    \begin{subfigure}[b]{0.5\textwidth}
        \centering
        \includegraphics[width=\textwidth]{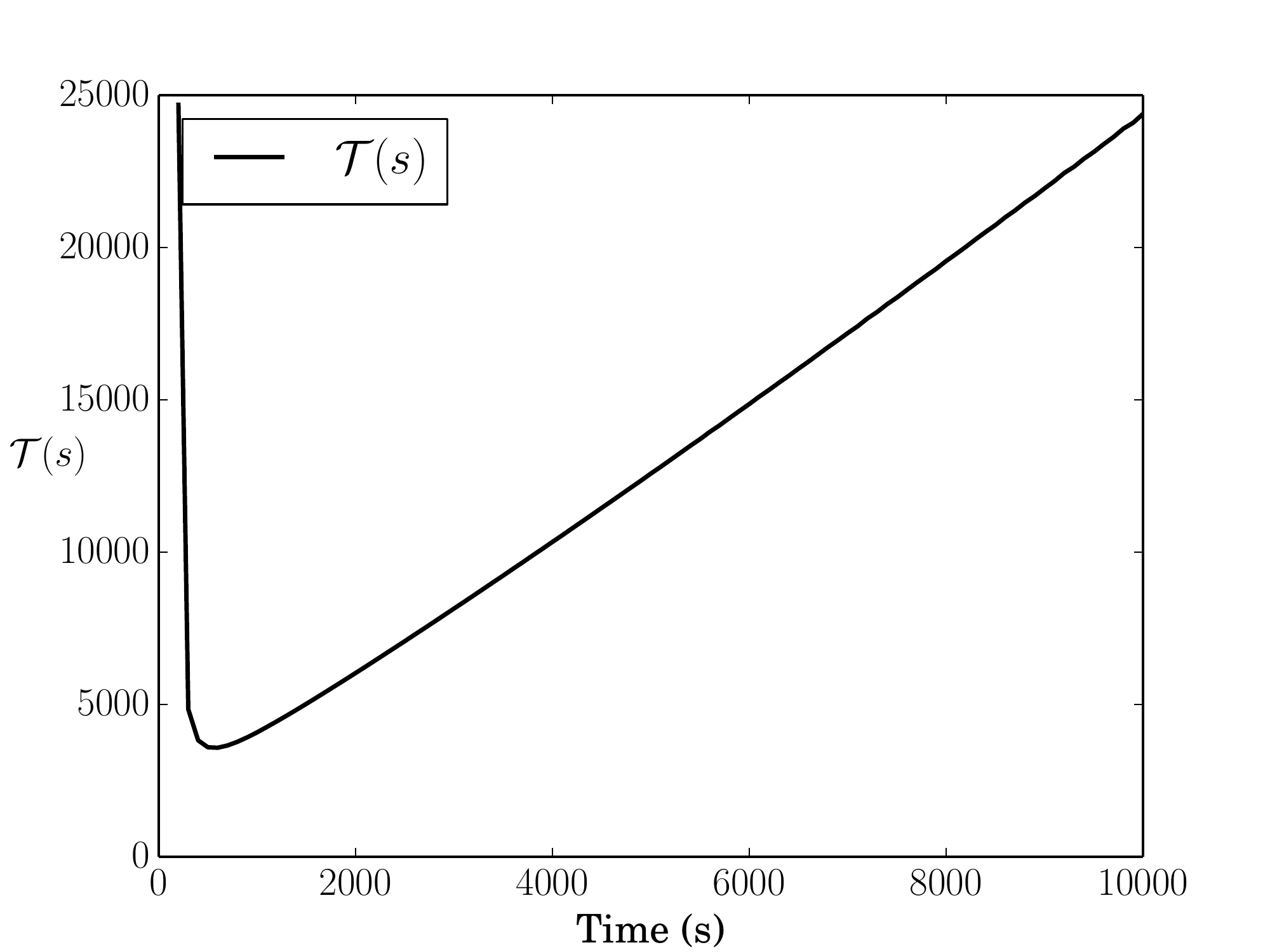}
        %\caption{Characteristic time $\mathcal{T}(s)$}
    \end{subfigure}
    \caption{Left: $C_\vac$ and its derivative as functions of time. Right: characteristic time $\mathcal{T}$ as a function of time.}
    \label{fig:TdCv}
\end{figure*}

After the nucleation stage as the system enters the growth regime, we use the time step $\Delta t^M = 10^{-3}$~s. This time step ensures a good conservation of the total quantity of matter $\mathcal{Q}_\mathrm{tot} = C_\vac + \sum_{n=2}^{N_\mathrm{max}} n C(n)$, with a relative error of order $10^{-7}$. The final time of computation $t_f$ is set to $10^4$ s. We denote by $C^\mathrm{ref}$ the numerical solution of the dynamics~\eqref{eq:EPn}--\eqref{eq:EPv} obtained by this procedure. The solution $C^\mathrm{ref}$ allows to compute the characteristic time $\mathcal{T}$ defined in~\eqref{eq:defT}. We observe in Figure~\ref{fig:TdCv} that $C_\vac$ strongly decreases at first from its initial value $C_\mathrm{init} = 10^{-7}$ and then enters a quasi-stationary state where the concentration of vacancy slowly decreases. At time $t = 10^{3}$ s, the characteristic time $\mathcal{T}(t)$ is approximately equal to $4\times 10^{3}$ s. This indicates that a time step $\delta t$ of order $10^{2}$ s is sufficiently small in order to keep the variations of $C_\vac$ small.\\

\begin{figure*}[htb]
    \centering
    \begin{subfigure}[b]{0.5\textwidth}
        \centering
        \includegraphics[width=\textwidth]{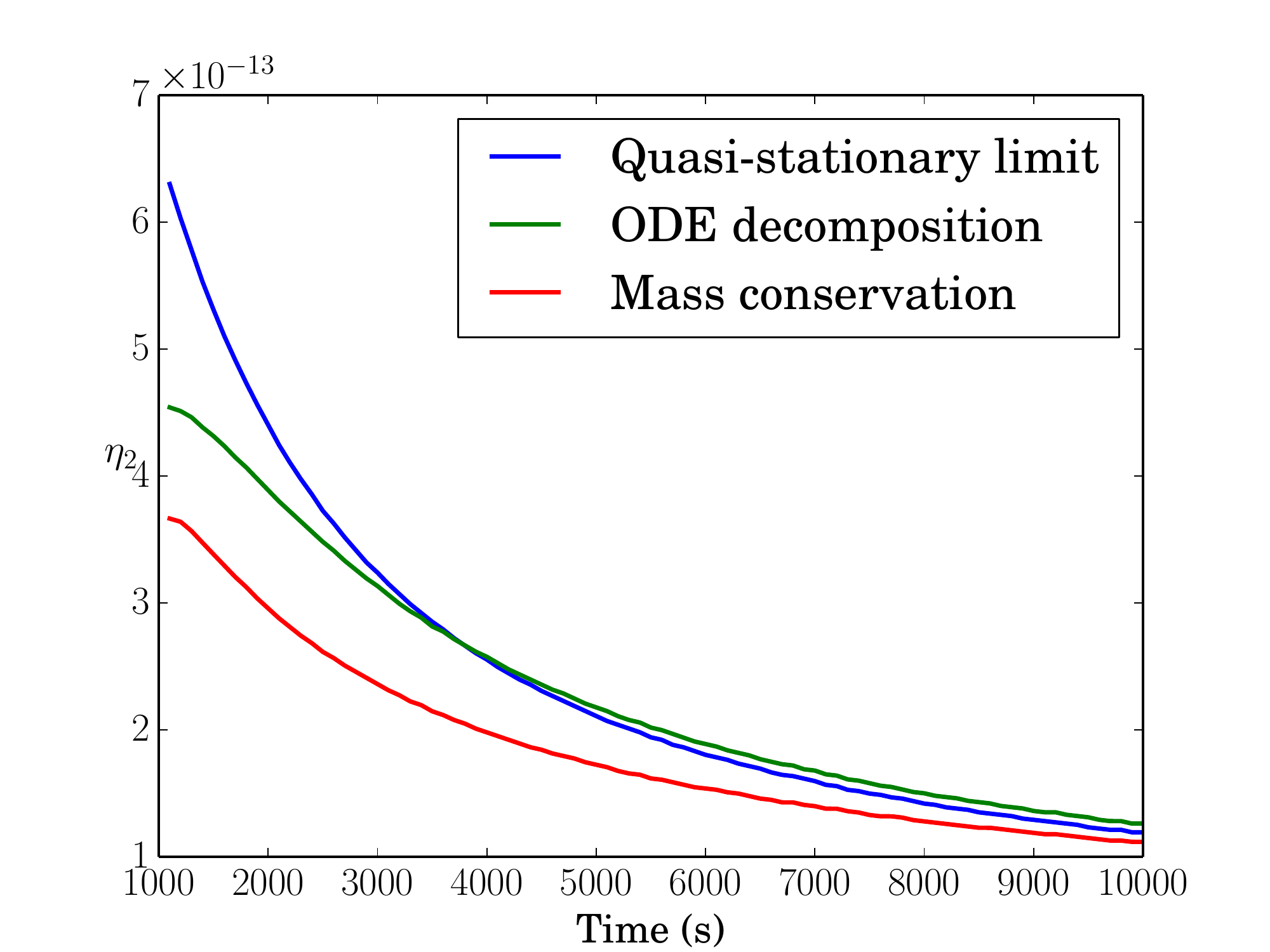}
        \caption{$\eta_2$ as a function of time}
    \end{subfigure}%
    ~
    \begin{subfigure}[b]{0.5\textwidth}
        \centering
        \includegraphics[width=\textwidth]{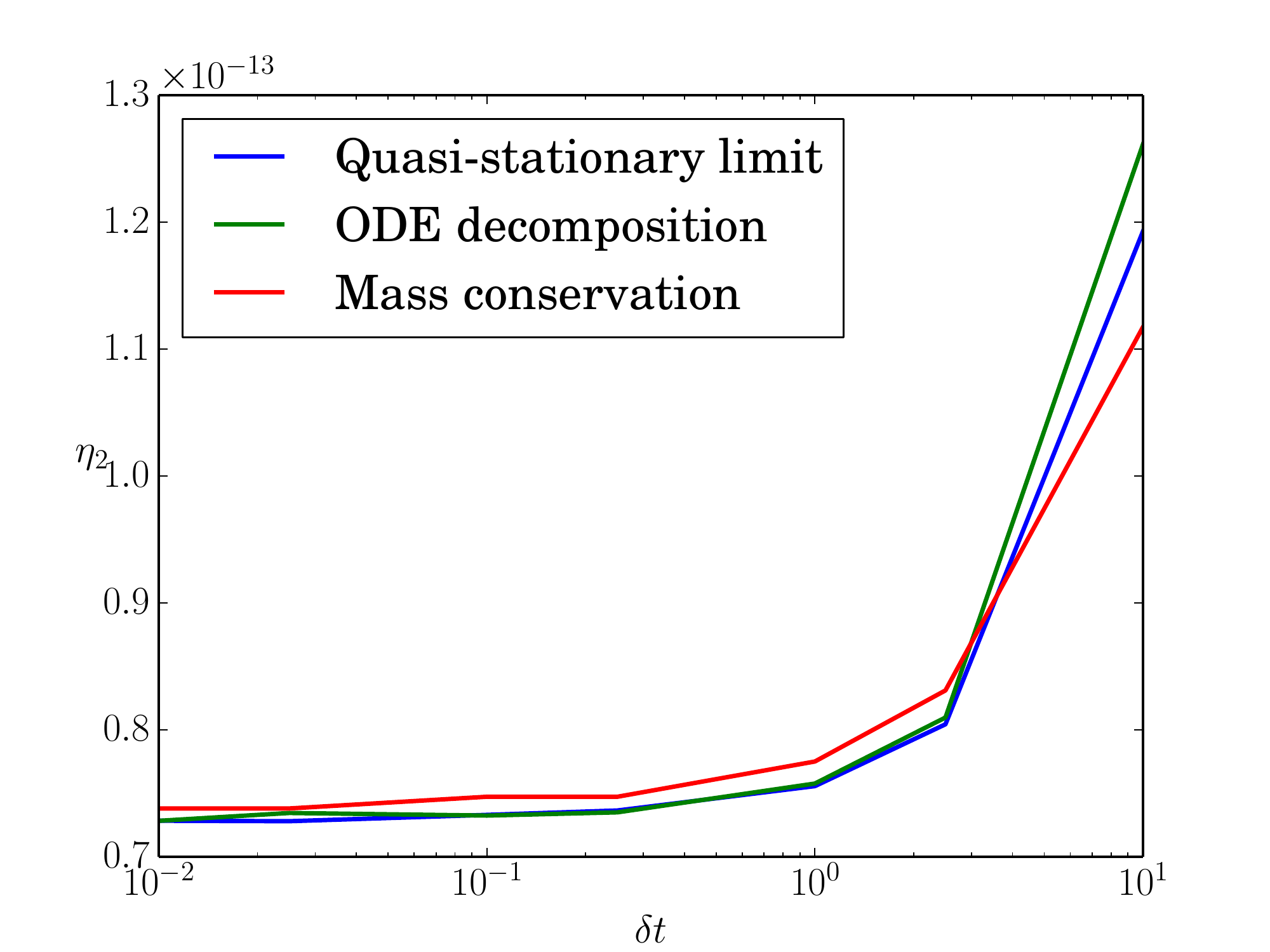}
        \caption{$\eta_2$ as a function of $\delta t$}
    \end{subfigure}
    \caption{Behaviour of the $\ell^2$-error}
    \label{fig:ErrorL2}
\end{figure*}

We next compare the three methods of computing $C_\vac$, using a Euler-Heun integration of problem~\eqref{P1} to update the cluster concentrations (still with $\Delta t^M = 10^{-3}$~s). The initial condition is set to $C^\mathrm{ref}(t_0)$ and $t_0 = 10^{3}$ s ensures that the initial transient regime is over. We then compute the solution obtained when updating the vacancy concentration every time step $\delta t = 10$ s using one of the three methods discussed in Section~\ref{sec:Cvac}, until $t_f = 10^{4}$~s. For the method presented in Section~\ref{sec:CvacODE}, we use $\Delta t = 10^{-3}$~s. We then estimate the $\ell^2$-error between the full ODE solution $C^\mathrm{ref}$ and the solutions $C^\mathrm{splitting}$ of problem~\eqref{P1}:
\begin{equation}
\eta_2(t) = \sqrt{\sum_{n=1}^{N_\mathrm{max}} \left( C^\mathrm{splitting}_n(t) -C^\mathrm{ref}_n(t)\right)^2}. 
\end{equation}
Figure~\ref{fig:ErrorL2}.a compares the errors for each of the three ways of integrating the dynamics of $C_\vac$ for $C^\mathrm{splitting}$. Figure~\ref{fig:ErrorL2}.b compares the error at the final time $t_f$, for various $\delta t$. There is no significant difference between the three methods. It seems that the $\ell^2$-error slightly decreases over time but increases exponentially with $\delta t$. Nevertheless the error remains more than 5 orders of magnitude lower than the total quantity of matter.

In the sequel, we choose the quasi-stationary limit approximation~\eqref{eq:QuasiStat} as it is stable and straightforward to compute. The conclusion could be different when other types of mobile clusters are taken into account (typically small clusters such as $C_2,\cdots,C_{10}$). In this case the mass conservation method cannot be used. The decomposition into elementary integrable ODEs then becomes the best alternative since the quasi-stationary limit approach requires solving a system of coupled second order equations.

\subsection{Accuracy of the splitting algorithm for thermal ageing}

We now present simulation results obtained with the main algorithm presented in Section~\ref{sec:algo}, using the methods presented in Section~\ref{sec:BD} and~\ref{sec:FP} for the large cluster dynamics. In the Steps~\eqref{algo:B2.a}--\eqref{algo:B2.b} and~\eqref{algo:L2} of the large size clusters dynamics, each particle is propagated independently, which allows dispatching the computations on a parallel architecture. The computation reported here were performed on a cluster of $15$ hyperthreaded cores, each thread being used to propagate $N_\mathrm{proc} = 2.10^5$ particles, which gives us a total of $N_\mathrm{sim} = 6.10^6$ particles. The final time of computation $t_f$ is set to $10^5$ s. %As explained in Section~\ref{sec:resampling}, the computation of $\widetilde{\mathscr{C}}$ at each time step $\delta t$ is time consuming, and depending of the number of particles $N_\mathrm{sim}$ and the time step $\delta t$, the computational time is reduced at least by a factor 10 or more if the number of particles increases or the time step decreases.\\

\begin{figure}[h!]
\centering
\includegraphics[width=0.5\textwidth]{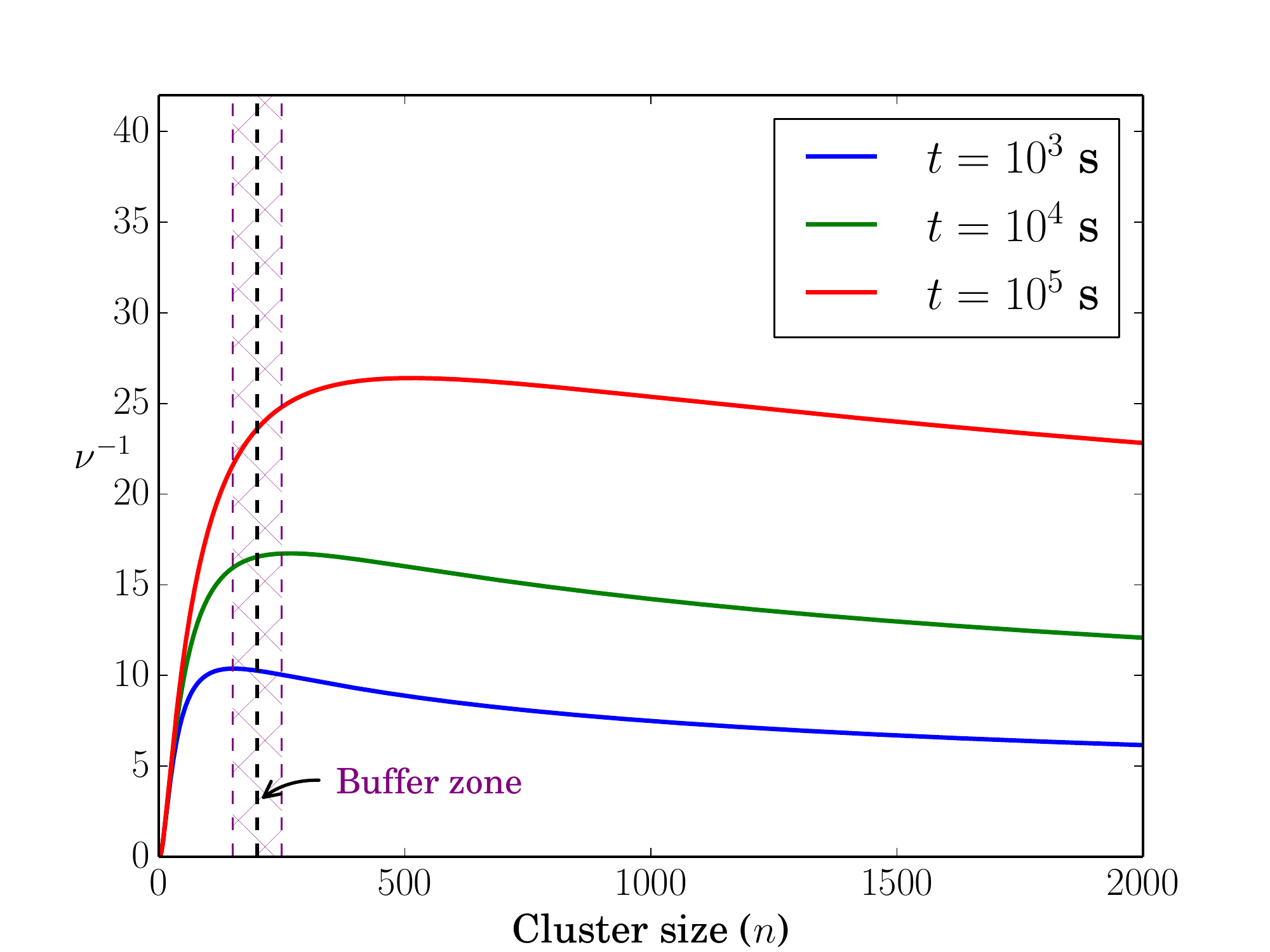}
\caption{Characteristic jump time $\nu^{-1}$ (see~\eqref{eq:nu}) as a function of the cluster size $n$: The highest frequent events occur for small $n$.}
\label{fig:nu}
\end{figure}

The time step used in Fokker-Planck simulations is set to $\Delta t^L = 1$ s while the concentration of vacancies $C_\vac$ is updated\footnote{We could have chosen larger time steps and time intervals in order to speed up the computational time. However we refrained from optimizing the parameters and comparing with state of the art methods since our main objective is to use our method in more complex problems than the ones which can be currently solved with classical methods.} at times that are multiple of $\delta t = 10$ s. The value of $C_\vac$ is calculated using the quasi-stationary limit approach~\eqref{eq:QuasiStat}. For the Birth-Death approach, the time step is not fixed but a characteristic jump time is given by the particles of size $N_\mathrm{front}$ and is of the order $(\beta_{N_\mathrm{front}}C_\vac + \alpha_{N_\mathrm{front}})^{-1}$ which is of order of $10$~s when the system is in a growth regime (see Figure~\ref{fig:nu}). Moreover it is increasing with time. In contrast, with a fully stochastic approach, the time steps of the most frequent events are of order $(\beta_{2}C_\vac + \alpha_{2})^{-1} \simeq 10^{-2}$ s, while the standard SSA approach requires us to choose one event at a time with characteristic time of order $10^{-11}$~s for boxes of volume $V = 10^{-18}$~m$^3$. \\

\begin{figure}[h!]
\centering
\includegraphics[width=0.5\textwidth]{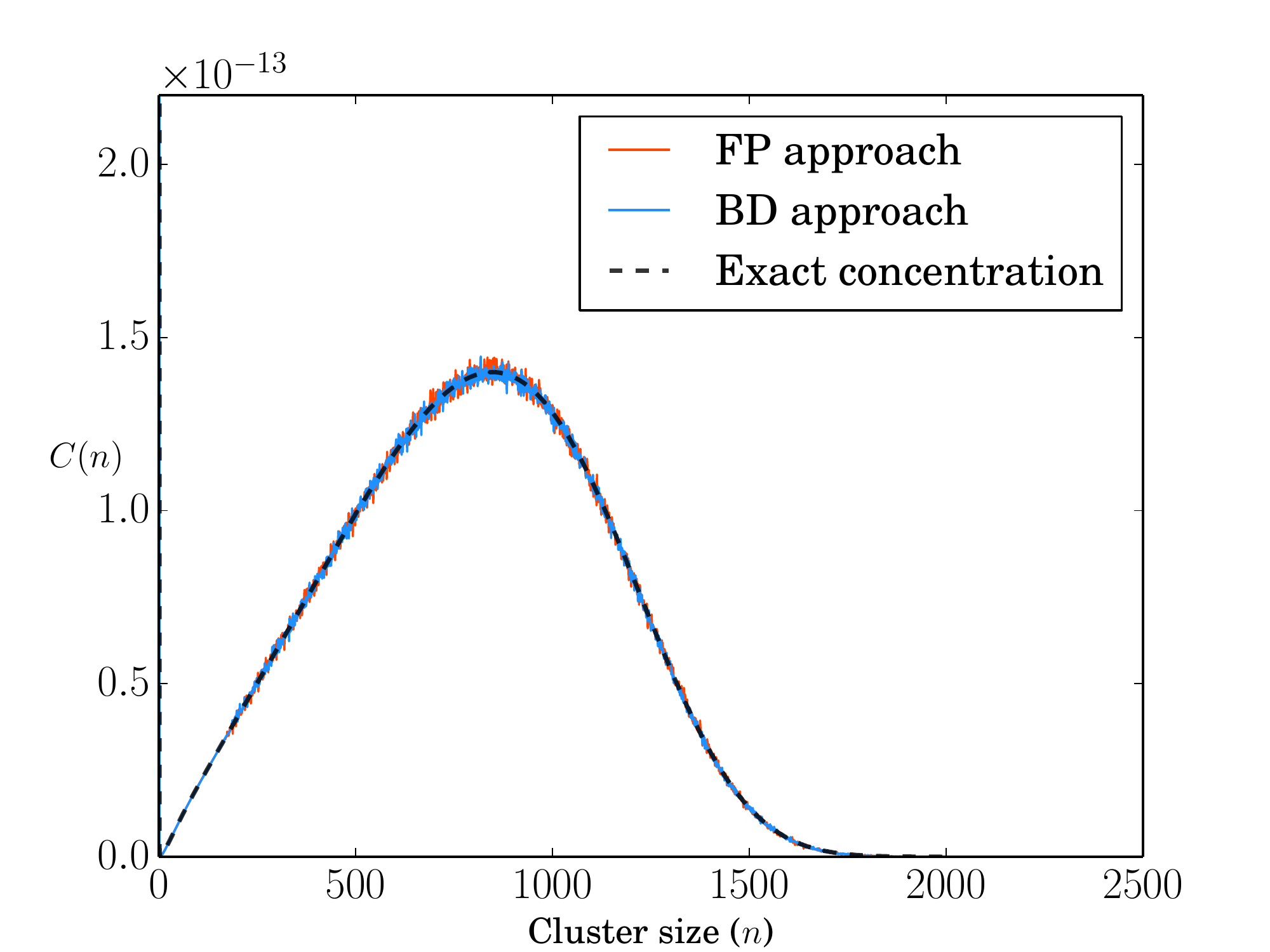}
\caption{Comparison between the exact concentration (black dotted line), the concentration obtained with a Fokker-Planck approach (red line) and the one obtained with the Birth-Death process approach (blue line).}
\label{fig:Error}
\end{figure}

Aside from the stochastic fluctuations inherent to both methods, the results presented in Figure~\ref{fig:Error} show a perfect agreement with the exact concentration obtained by an integration of the full ODE system. The total concentration is equal to $\mathcal{Q}_\mathrm{tot}^{\rm FP} = 9.989\times 10^{-8}$ at the final time in the case of the FP approach and is equal to $\mathcal{Q}_\mathrm{tot}^{\rm BD} = 9.968\times 10^{-8}$ for the BD approach. The relative error on the total concentration is therefore less than $0.4 \%$.

\section{Conclusion}
\label{sec:conclusion}
In this article we have presented a generic coupling algorithm allowing to simulate thermal ageing and ageing under irradiation using cluster dynamics. Our approach consists in coupling the standard rate equations for small size clusters with more efficient methods for large size clusters. Such a coupling is based on a splitting of the dynamics between the nonlinear dynamics of the vacancy concentration and the linear evolution of the cluster concentrations at fixed vacancy concentration. The dynamics of cluster concentrations is integrated by decomposing the initial condition and independently evolving the dynamics of small and large clusters.\\

We emphasized two stochastic methods in order to simulate the evolution of the concentration of large size clusters. The Fokker-Planck approach is well known, but our stochastic treatment with Langevin dynamics is a recent approach~\cite{surh2004master} in the cluster dynamics community. The Birth-Death process approach is reminiscent of the SSA algorithm but can be parallelized much more efficiently and we avoid high frequency events associated with small size clusters. The main interest of these approaches is that it can be extended to higher dimensional situations. Moreover, with both methods, the particles are propagated independently, which allows to dispatch computations on a parallel architecture, henceforth decreasing the wall-clock computation time. With both methods the quantity of matter is accurately conserved and the distribution of concentrations we obtain is very close to the exact solution obtained by a numerical integration of the original full ODE system.\\

The paper is primarily intended to introduce a new method in the treatment of large scale problems in cluster dynamics which might be difficult to solve with existing methods. In a future work we plan to present the effectiveness of such hybrid methods to solve problems with two or more types of species.

%\section*{Acknowledgements}

%\appendix

%\section{Section in Appendix}
%\label{appendix-sec1}

%Sample text. Sample text. Sample text. Sample text. Sample text. Sample text. 
%Sample text. Sample text. Sample text. Sample text. Sample text. Sample text. 
%Sample text. 

%% References


\bibliographystyle{elsarticle-num}

\bibliography{sample}

\end{document}